\documentclass[a4paper]{jpconf}
\usepackage{graphicx}
\usepackage[printwatermark]{xwatermark}
\usepackage{xcolor}
\usepackage{upgreek}
\usepackage{siunitx}
\usepackage{svg}
\usepackage{textcomp}

\newsavebox\mybox
\newwatermark*[
  allpages,
  angle=70,
  scale=15,
  xpos=-90,
  ypos=15
]{\usebox\mybox}
\definecolor{prlblue}{rgb}{0.176, 0.152, 0.57} 
\usepackage[colorlinks=true, linkcolor=black, citecolor=prlblue, filecolor=black, urlcolor=prlblue]{hyperref} 

\begin{document}

\title{Tunable and precise two-bunch generation at FLASHForward}

\author{S~Schr\"oder$^{1,2}$, K Ludwig$^1$, A Aschikhin$^{1,2}$, \mbox{R D'Arcy$^1$}, M Dinter$^1$, P Gonzalez$^{1,2}$, \mbox{S Karstensen$^1$}, A Knetsch$^1$, V Libov$^{1,2}$, C A Lindstr{\o}m$^1$, F Marutzky$^1$, P Niknejadi$^1$, A Rahali$^1$, L Schaper$^1$, \mbox{A Schleiermacher$^1$}, \mbox{B Schmidt$^1$}, \mbox{S Thiele$^1$}, A de Zubiaurre Wagner$^1$, S Wesch$^1$ and J Osterhoff$^1$}
\vspace{0.7cm}
\address{$^1$ Deutsches-Elektronen Synchrotron (DESY), Notkestra{\ss}e 85, 22607 Hamburg, Germany}
\address{$^2$ University of Hamburg, Luruper Chaussee 149, 22761 Hamburg, Germany}

\ead{sarah.schroeder@desy.de}

\begin{abstract}
Beam-driven plasma-wakefield acceleration based on external injection has the potential to significantly reduce the size of future accelerators. Stability and quality of the acceleration process substantially depends on the incoming bunch parameters. Precise control of the current profile is essential for optimising energy-transfer efficiency and preserving energy spread. At the FLASHForward facility, driver--witness bunch pairs of adjustable bunch length and separation are generated by a set of collimators in a dispersive section, which enables \mbox{fs-level} control of the longitudinal bunch profile. The design of the collimator apparatus and its commissioning is presented.
\end{abstract}

\section{Introduction}
High acceleration gradients inherent in a plasma wakefield make it a compelling technique for compact particle accelerators \cite{TajimaDawson1979, Ruth1984, Chen1985}, potentially reducing costs of future free-electron lasers (FELs) or colliders \cite{ CBSchroeder2010, Adli2013, CBSchroeder2016}. For such applications, beam-driven plasma-wakefield acceleration (PWFA) is a promising approach, as it can supply high wall-plug efficiencies and MW-scale average power. Studies on the external-injection scheme, where the energy is transferred from a driver to a witness bunch, are crucial for staging plasma channels to achieve the required energies (GeV--TeV). High-gradient as well as high-efficiency acceleration using the external-injection PWFA method have successfully been demonstrated \cite{Blumenfeld2007, Litos2014}.

The next milestone of external-injection PWFA is high-efficiency acceleration while preserving the beam quality, including emittance and energy spread. The energy-transfer efficiency contributes largely to the overall efficiency of the acceleration process \cite{Ruth1985, Chen1986, Chiou1998, Lotov2005}. For high-energy applications, also the transformer ratio---the limitation on the energy being transferred from the wakefield-driving bunch to the accelerated bunch---must be optimised. Asymmetric longitudinal bunch profiles can lead to an increased transformer ratio \cite{Loisch2018}. Furthermore, the resulting energy spread of the accelerated bunch is determined by the wakefield structure, which also changes with the current profiles of the driver and witness bunches. Current profile shaping of the incoming bunches can thus achieve high efficiency as well as energy-spread preservation via optimised beam loading \cite{Katsouleas1987}.\\

\begin{figure}
    \centering
    \includegraphics[width=\textwidth]{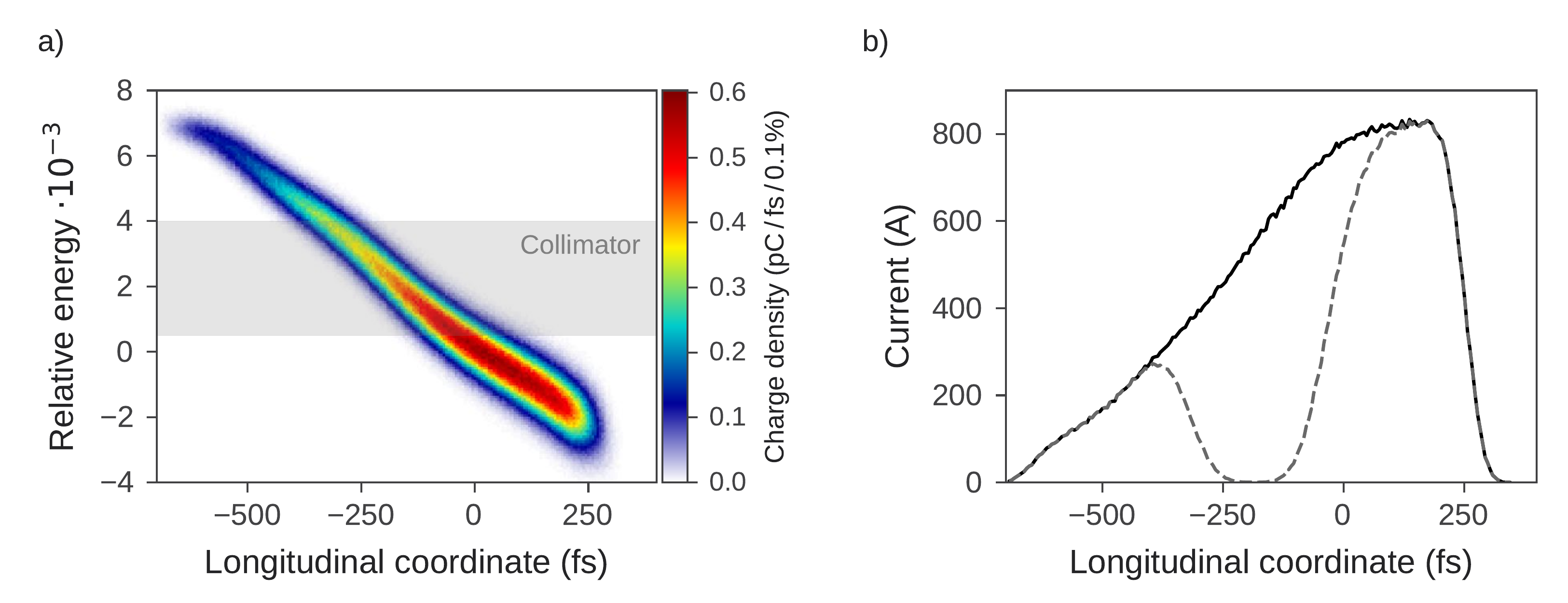}
    \caption{\textbf{Two-bunch generation using a collimator in a dispersive section.} The longitudinal phase space is linearised in the FLASH accelerator and measured by a transverse deflecting structure in a photon beamline parallel to FLASHForward (a). A collimator in a dispersive section effectively cuts the original bunch (solid) in a driver--witness bunch pair (dashed) (b).}
    \label{fig:LPS}
\end{figure}

Several different techniques for longitudinal bunch shaping have already been explored. Laser pulse stacking at the photo-cathode \cite{Will2008, FerrarioM2011, RonsivalleC2014, PompiliR2016, LoischNIM2018} as well as emittance exchange \cite{PEmma2006, PPiot2011, QGao2018} can be used to tailor the current profile of the electron bunches as it is generated. Both techniques have independent control of the driver and witness current profiles. Partial beam masking with a collimator is also a well established method for two-bunch generation \cite{Muggli2008, Litos2014}. Owing to a strong energy--time correlation of the electron bunch, a collimator in a dispersive section acts effectively also on the longitudinal bunch profile (see Figure \ref{fig:LPS}b). Thus, this method relies on longitudinal-phase-space shaping of a chirped bunch in a pre-accelerator.

At future high-energy accelerators, the drive bunch must be generated independently from the witness bunch before each plasma stage. The collimation two-bunch generation method can thus only be applied for single plasma channels. Nevertheless, the intrinsic synchronisation of the two bunches and the simple implementation to pre-existing beamlines make it a convenient option for underlying single-stage studies and is, as such, also implemented at FLASHForward.\\

FLASHForward is an experimental test facility dedicated to beam-driven plasma-wakefield research \cite{Darcy2019} and is located in an extension beamline to the X-ray Free-electron LASer in Hamburg (FLASH) at DESY \cite{Ayvazyan2006, Ackermann2007}. Primary scientific goals are high-brightness electron-bunch generation from the plasma (X-1: Internal injection) and beam-quality-preserving acceleration of pre-existing bunches (X-2: External injection). Furthermore, the access to \SI{3}{MHz} micro-pulse repetition rate makes FLASHForward a unique facility for high-average-power (up to \SI{30}{kW}) studies on beam-driven PWFA (X-3: High average power).

\section{Longitudinal phase-space shaping at FLASHForward}
In the FLASH accelerator, electron bunches of up to \SI{1}{nC} are produced at \SI{10}{\hertz} in a laser-driven photoinjector and subsequently accelerated up to \SI{1.2}{GeV} in seven \SI{1.3}{\giga \hertz} superconducting radio-frequency (SRF) cavities. Two magnetic compressor chicanes can reduce the bunch length to \SI{150}{fs} (rms) with peak currents on the order of 1-2\,kA.

A series of SRF cavities and bunch compressors in the FLASH linear accelerator allows for flexible longitudinal phase-space shaping \cite{Piot2012}. Off-crest acceleration in the SRF cavities imprints a correlation between the particle energy and the longitudinal bunch coordinate---an energy chirp. The chirped bunch is then longitudinally compressed in the magnetic chicanes of FLASH and in the FLASHForward extraction line. A third-harmonic SRF cavity in the FLASH accelerator gives control over the first and second derivative of the chirp. This allows in particular linearisation of the longitudinal phase space (see Figure \ref{fig:LPS}a).\\

The generation of driver-witness bunch pairs for external-injection PWFA is not directly supplied as a standard operation mode at FLASH. The two-bunch generation for FLASHForward can either be realised at the gun by two delayed laser pulses or by a set of collimators in a dispersive section blocking the middle part of a single bunch.  This paper reports on the two-bunch generation by collimation, which in particular provides high tunability and precision within the restriction of the original beam profile.

\section{Collimator design}
\begin{figure}
\noindent
\begin{minipage}[b]{.55\textwidth}
  \centering
  \includegraphics[clip,trim=6.5cm 1cm 0cm 8,width=\linewidth]{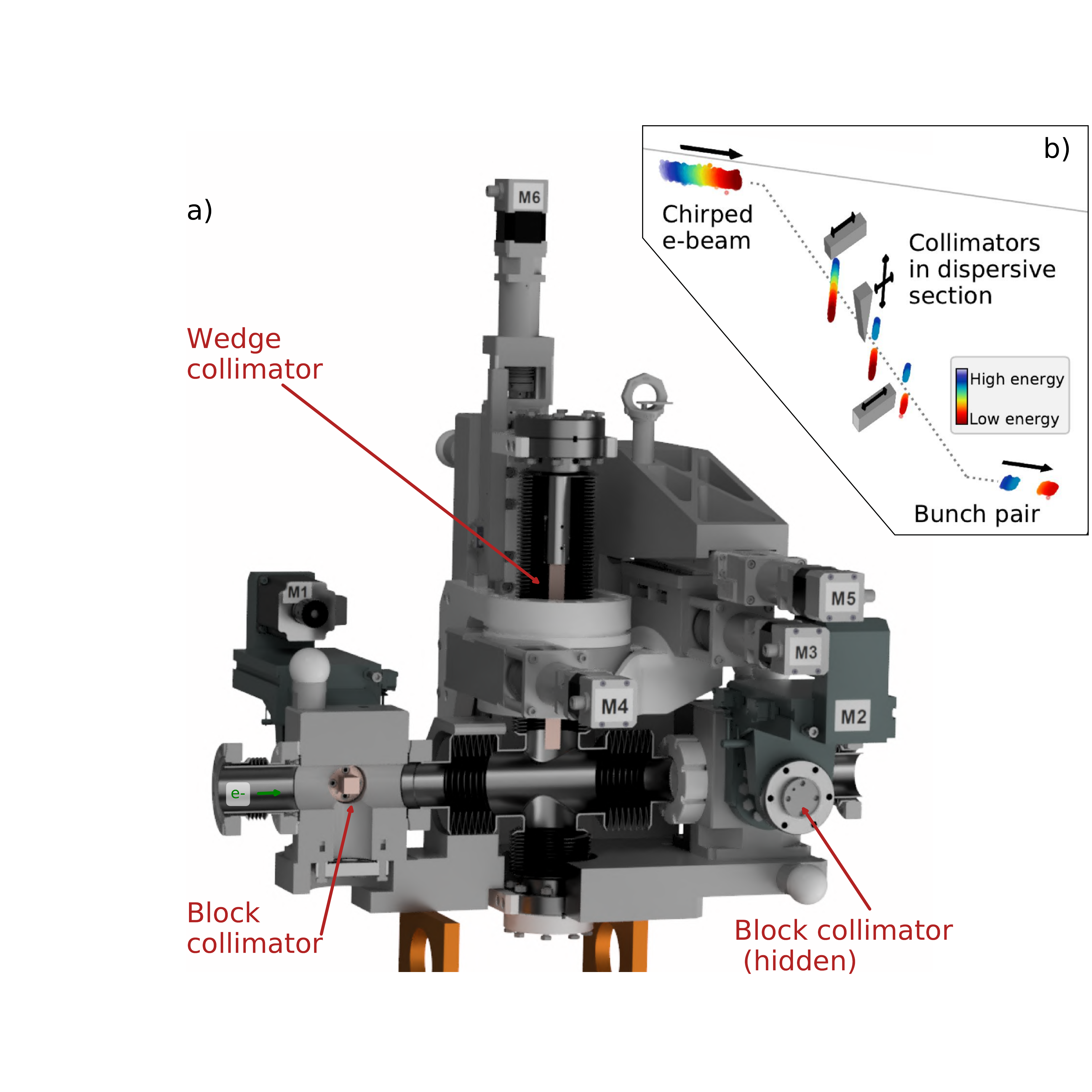}
\end{minipage}%
\hfill%
\begin{minipage}[b]{.4\textwidth}
  \centering
  \includegraphics[width=\linewidth]{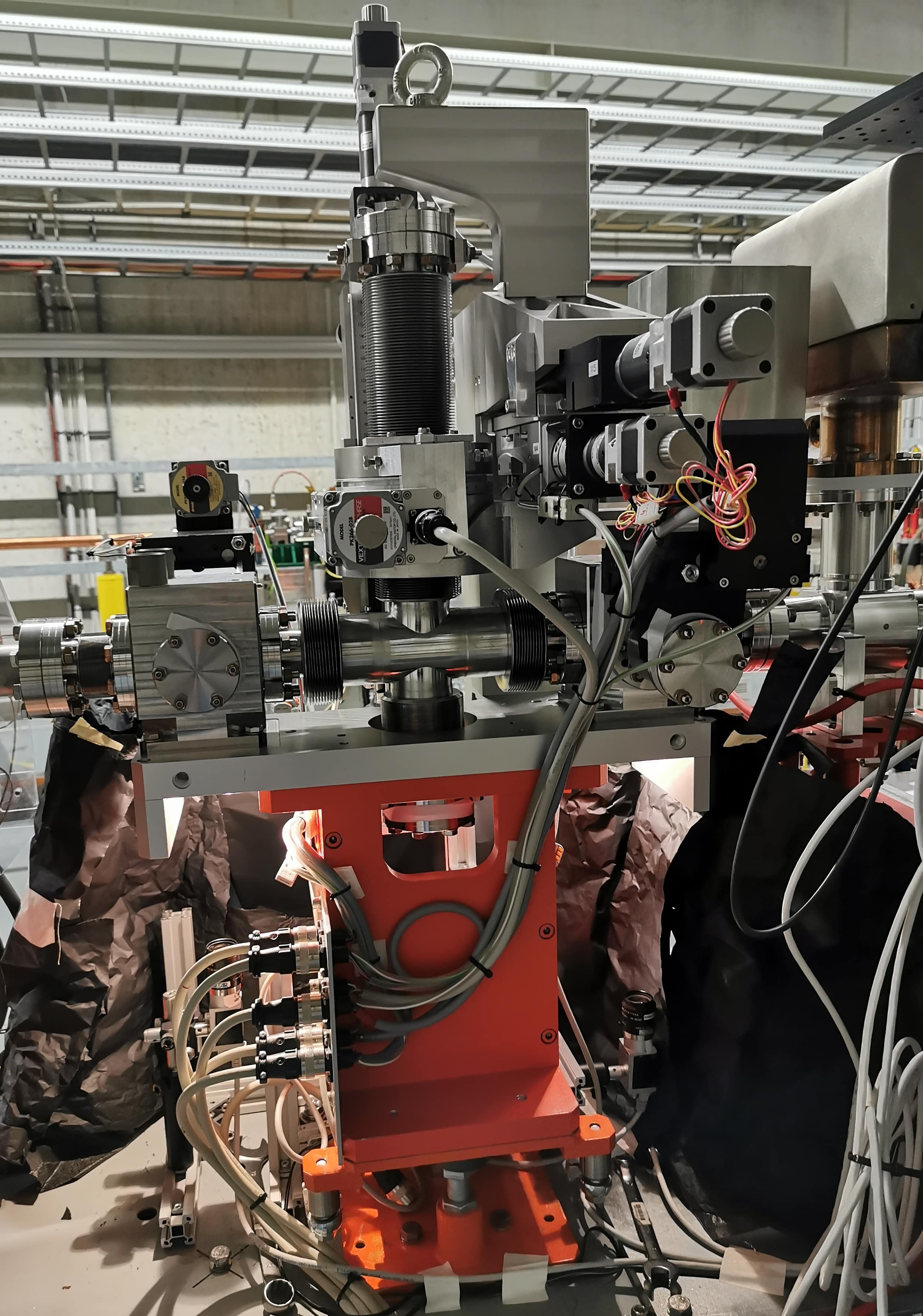}
\end{minipage}
\par
\begin{minipage}[t]{.55\textwidth}
  \raggedleft
  \caption{\textbf{3D CAD model of the three collimator device.} (a) The block collimators are adjustable in the horizontal plane (M1, M2) and act on the head and tail of the bunch. The wedge determines the bunch separation width (M6) and position (M5). Further alignment of the wedge is available by rotation about the vertical (M4) and the longitudinal (beam) axis (M3). (b) Schematic layout of bunch collimation in dispersive section.}
  \label{fig:ScraperCAD}
\end{minipage}%
\hfill%
\begin{minipage}[t]{.4\textwidth}  
  \raggedright 
  \flushright
  \caption{\textbf{Installed collimator device.} An adjustable support allows precision alignment to the beamline. Cameras observe the collimator position through window flanges from underneath. All collimators can be adjusted remotely.}
  \label{fig:ScraperImage}  
\end{minipage}  
\end{figure}

At FLASHForward, a three-collimator device is implemented (see Figure \ref{fig:ScraperCAD} and Figure \ref{fig:ScraperImage}), which allows the beam to be manipulated in four distinct and independent ways: separation width, separation position, driver and witness length.\\

The width and position of the driver--witness separation is determined by a wedge-shaped collimator that can be adjusted by stepper motors vertically (M6) and horizontally (M5). The vertical movement determines the width of the collimator and thus the separation width. The horizontal movement determines the separation position and thus the charge distribution and bunch lengths. For alignment, the wedge collimator can additionally be rotated about the vertical (M4) and the longitudinal (beam) axis (M3).

The required wedge geometry is given by the typical energy spread   (0.1--0.5\% rms) of the incoming electron bunch and the horizontal dispersion at the location of the collimators (\SI{-340}{mm}). This results in a transverse bunch size of $1.5 \times 0.1$~mm rms (x $\times$ y). At the plasma channel the bunches typically have a bunch length of \SIrange{150}{300}{fs} rms, which is suitable for external injection experiments at plasma densities between $10^{15}$--$10^{17}~\mathrm{cm^{-3}}$.

The wedge-shape allows stepless and precise separation-width adjustment. The available width of \SIrange{0.6}{3}{mm} ensures considerable cutting of the bunch, resulting in \SIrange{30}{120}{\um} bunch separations. The wedge height of \SI{125}{mm} makes the vertical cutting across the bunch negligible (below 1\%). The depth of \SI{15}{mm} ensures sufficient particle scattering, structure stability and limited energy deposition in the material. The multi-dimensional tension from wedge adjustments (two linear axles, a goniometer, and a circle segment) is decoupled from the beam pipe with a cross-shaped bellow construction.\\

In front of and behind the wedge collimator two block-shaped collimators are installed, which are adjustable in the dispersive (horizontal) plane. For the standard FLASH operation mode with a negatively chirped bunch, the upstream collimator acts on high energies---the bunch tail (M1)---and the downstream collimator acts on low energies---the bunch head (M2). The block collimators are produced in a simple rectangular solid shape (\num[output-product = \times]{10 x 10 x 15 }{ mm}). This geometry also allows for blocking the entire bunch.

These block collimators also have the capability of individual characterisation of the drive and witness bunch by blocking one or the other bunch. Furthermore, in case of a sufficiently small slice energy spread, the simultaneous movement of the block collimators---so that only a small energy range can pass---enables \textit{temporally sliced bunch characterisation}.\\

\begin{table}[ht]
\begin{center}
\begin{tabular}{ l l l l l }
\br
Collimator & Motor & Sledge type, DOF & Range & Accuracy \\
\br
Block & M1  & lin. axle,  hor. translation & \SI{30}{mm} & \SI{10}{\um} \\
\mr
Block & M2  & lin. axle,  hor. translation & \SI{30}{mm}  & \SI{10}{\um} \\
\mr
Wedge & M3 & Circle segment,  long. rotation & $\pm 1^{\circ}$ & $0.001^{\circ}$\\
      & M4 & Goniometer,  vert. rotation & $\pm 1^{\circ}$ & $0.001^{\circ}$\\
      & M5 & linear axle,  hor. translation & $\pm 3$ mm & \SI{20}{\um} \\
      & M6 & linear axle,  vert. translation & \SI{120}{mm} & \SI{20}{\um} \\
\br
\end{tabular}
\caption{ \textbf{Technical specification of collimators.} Six remotely controllable motors are available for tunable two-bunch generation. The $\upmu$m-precision control of the linear feedthroughs ensure the required fs-precision of bunch lengths.}
\label{tab:Motors}
\end{center}
\end{table}

\clearpage

An emphasis was placed on \textit{beam loading control} and the capability of \textit{acceleration optimisation precision studies} when designing the device. Start-to-end simulations \cite{Floettmann, Borland} suggest $\upmu$m-precision control of all collimators to control bunch lengths on the fs-level. Table \ref{tab:Motors} summarises the available collimator motors, their technical realisation and control accuracy.\\

All three collimators are made of copper-tungsten alloy (50:50). This material is compatible with ultra-high-vacuum operation and can be eroded to manufacture the sophisticated wedge geometry. Tungsten enhances the absorption, copper improves the heat transfer to the outside. A single shot \SI{1}{GeV} beam shows in Geant4 \cite{Allison} simulations 5\% and 15\% energy deposition, for a typical two-bunch separation at the wedge and a complete beam blocking with a block collimator, respectively. Heat generation in the collimators limits their use for  high-average-power studies.\\

The collimator apparatus approximately weighs \SI{120}{kg} and has a total beamline length of \SI{0.5}{m}. All collimator adjustments can be remotely controlled, with their positions observed by cameras through window flanges.

\section{ Commissioning of the collimators}
\begin{figure}[ht]
\includegraphics[width=\textwidth]{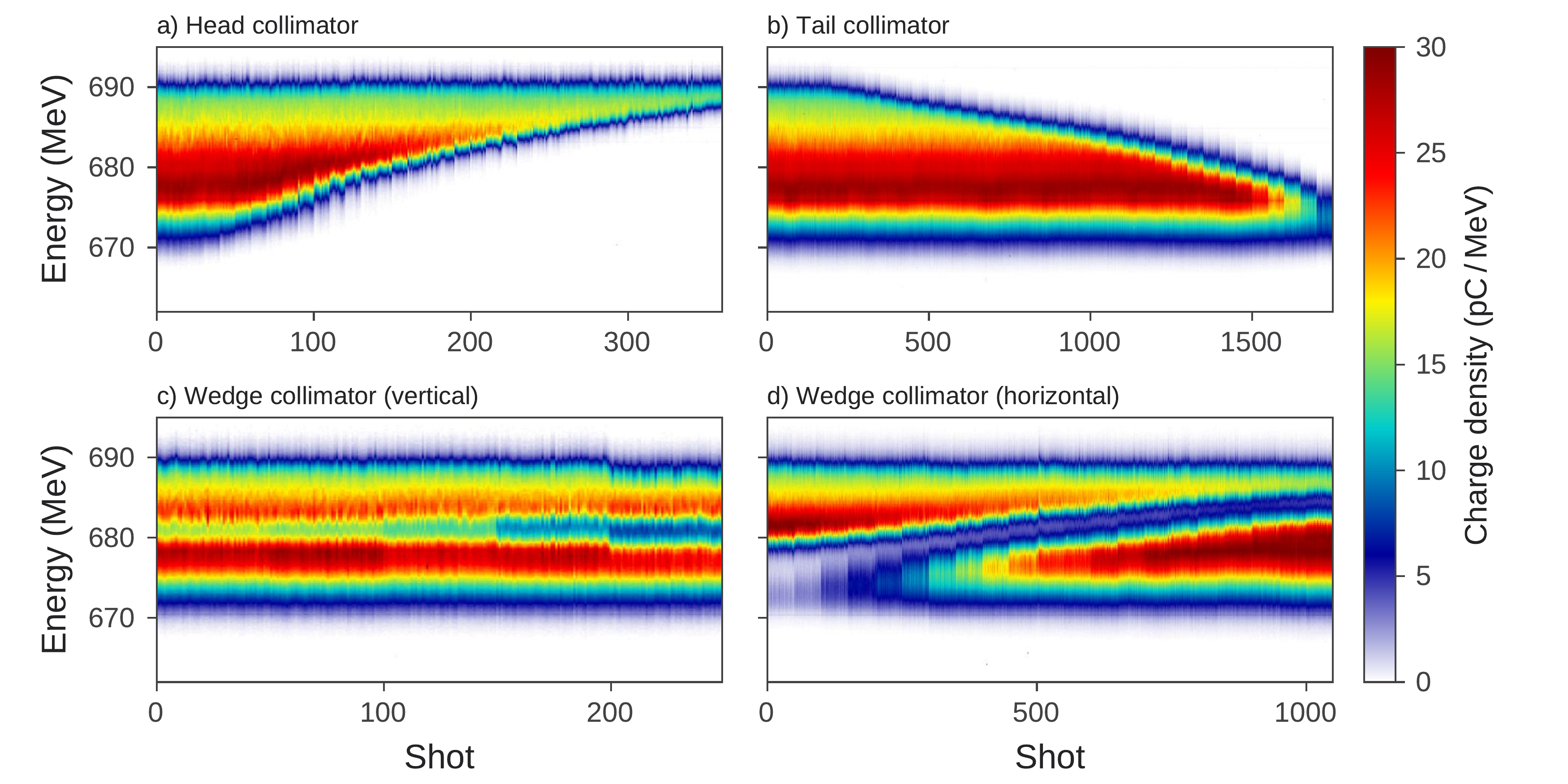}
\caption{\label{fig:MotorScans} \textbf{Energy projection waterfall of collimator position scans.} Two blocks can be used to remove the low-energy head (a) or the high-energy tail (b) of the bunch. Separation width (c) and position (d) is adjusted by a wedge-shaped collimator, which can be positioned in the vertical and horizontal (dispersive) plane.}
\end{figure}

The beam-based commissioning comprised of collimator position scans (see Figure \ref{fig:MotorScans}), which were recorded on a vertically-dispersive electron spectrometer. The horizontal collimator positions were scanned over the entire dimension of the bunch (Figure \ref{fig:MotorScans} a, b, d). Driver-witness bunch separations were recorded for 5 different vertical wedge positions (Figure \ref{fig:MotorScans} c). At each collimator position, about 20--50 consecutive shots were taken.\\

All collimator scans show a clean beam removal in the energy plane. The stability of the energy profiles at each collimator position relies on a high beam stability and especially a high pointing stability at the collimators. The combination of fine energy-profile tunability and a strongly correlated longitudinal phase space (see Figure \ref{fig:LPS}) allow precise two-bunch generation at FLASHForward.

\begin{figure}[ht]
\includegraphics[width=\textwidth]{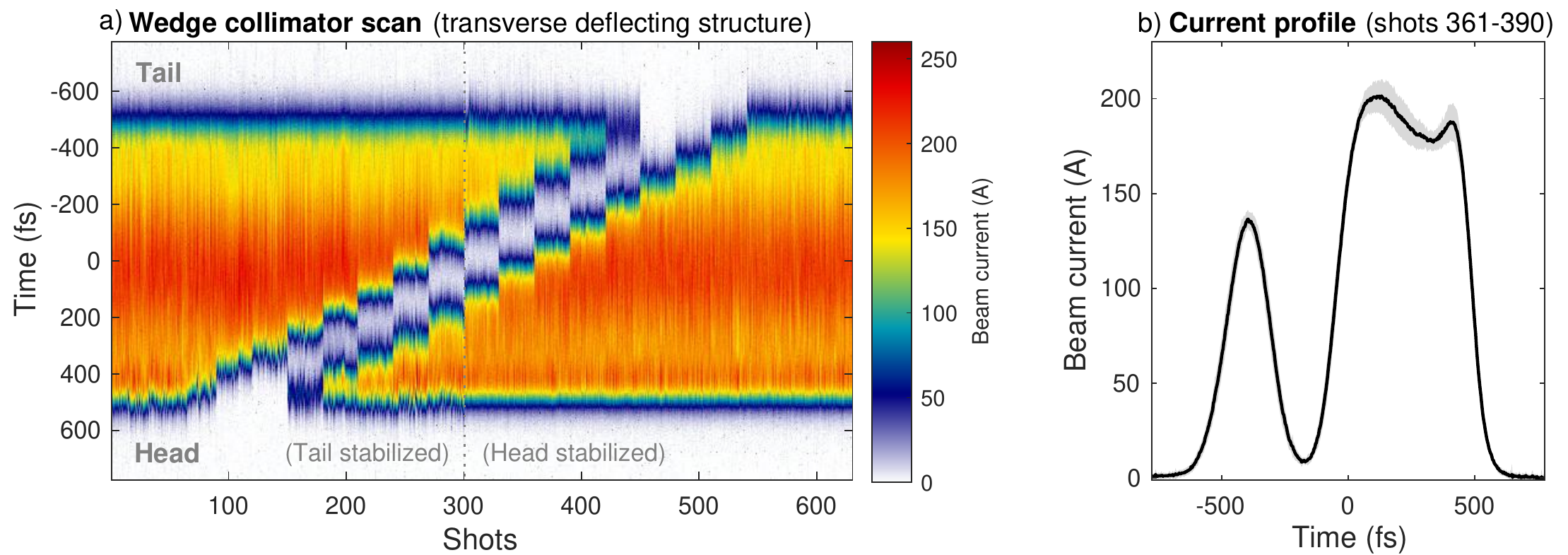}
\caption{\label{fig:MotorScansT} \textbf{Two-bunch measurement in a transverse deflecting structure.} Current profile waterfall of horizontal wedge position scan with a moderately compressed beam (a). The single shots are phase corrected with a threshold for the bunch tail (first $\sim$ 300 shots) and then for the bunch head. Mean current profile and its root mean square (gray band) of one particular collimator position (b).}
\end{figure}

The conceptual method could be verified with a recently commissioned transverse deflecting structure \cite{Pau2020} in the FLASHForward beamline (see Figure \ref{fig:MotorScansT}). The depicted stability of bunch separation width and separation position underpins the ability of adjustable and precise two-bunch generation at FLASHForward.

\section{Conclusion and outlook}
The presented two-bunch generation at FLASHForward with a set of collimators in a dispersive section demonstrate for high tunablity and flexibility of the driver--witness parameters. This method allows high-precision studies of the plasma acceleration as a function of the incoming bunch parameters.

The applicability of this method is limited by the slice energy spread of the original bunch, the available dispersion of the beamline, and the monotonicity of the energy chirp. Furthermore, the longitudinal-phase-space shape of the original bunch restricts the achievable bunch parameters. Consequently the longitudinal phase-space shape of the original bunch must be set up to the demands of the experiment.

First driver--witness bunch pairs for external injection PWFA at FLASHForward have been generated. Stability of the bunch charge and beam pointing at the collimator resulted in a repeatable and stable two-bunch structure, that is highly controllable. Precision studies of the plasma wakefield acceleration utilising the tunability and precision of the driver--witness bunch pair settings are ongoing.

\section*{Acknowledgement}
This work was supported by the Helmholtz ARD. The authors would like to thank E.-O. Saemann and the staff of DESY MVS group for their valuable engineering and technical support.

\end{document}